\documentclass[11pt,english]{article}
\usepackage{iftex}
\usepackage{amssymb,amsmath}
\ifPDFTeX
  \usepackage[T1]{fontenc}
  \usepackage[utf8]{inputenc}
  \usepackage{lmodern}
\else
  \usepackage{unicode-math}
  \defaultfontfeatures{Scale=MatchLowercase}
  \defaultfontfeatures[\rmfamily]{Ligatures=TeX,Scale=1}
\fi
\usepackage[main=english]{babel}
\usepackage{microtype}
\usepackage[margin=1in]{geometry}
\usepackage{enumitem}
\usepackage{graphicx}
\usepackage{float}
\usepackage{xurl}
\usepackage[unicode,hidelinks]{hyperref}
\usepackage{bookmark}
\hypersetup{
  pdftitle={Time-Travel Paradox in a Bohmian Extension of Deutsch's Approach: Mandatory Tunneling as the ``Banana Peel Mechanism''},
  pdflang={en},
  pdfcreator={LaTeX}}
\newcounter{assumption}
\renewcommand{\theassumption}{A\arabic{assumption}}
\newcommand{\assumption}[2]{%
  \par\refstepcounter{assumption}\label{#1}%
  (\theassumption) #2}

\title{Time-Travel Paradox in a Bohmian Extension of Deutsch's Approach:
Mandatory Tunneling as the ''Banana Peel Mechanism''}
\author{Vladimir Nikitin\thanks{riseway@proton.me}\\\small Independent Researcher}
\date{November 17, 2026}

\begin{document}
\maketitle

\begin{abstract}

We consider a self-prevention paradox in which a particle travels
through a wormhole into its own past, encounters its younger version,
and blocks its movement. In a specially constructed one-dimensional
model, this collision rules out a classically self-consistent history.

Deutsch's quantum approach \cite{deutsch-1991} can resolve such contradictions by imposing a
self-consistency condition on the density matrix. The physical
interpretation of this approach, however, remains debated. In
particular, Dunlap \cite{dunlap-2016} argues that Deutsch's model presupposes the
existence of interacting parallel worlds.

Here we extend Deutsch's approach within de Broglie--Bohm theory \cite{de-broglie-1928,bohm-1952} by
supplementing the self-consistency condition on the quantum state with a
self-consistency requirement for the particle's ontic configuration.
This allows us to consider the self-consistent evolution of an
individual particle. In a two-channel, narrow-band Gaussian
approximation, we find a self-consistent solution in which quantum
tunneling is mandatory, despite its arbitrarily small probability in an
ordinary scattering process. This realizes the ``banana peel mechanism'': to avoid a contradiction during a journey into the past, an
event that would ordinarily be extremely unlikely occurs nonetheless.

\end{abstract}


\section{Introduction}\label{introduction}

Some solutions of general relativity admit closed timelike curves,
although whether they can be physically realized remains an open
question \cite{godel-1949,luminet-2021,hawking-1992,morris-1988}. In particular, traversable wormholes have been
discussed as a possible basis for time machines, whereas the chronology
protection conjecture holds that the laws of physics prevent such curves
from forming. Without assuming that time machines actually exist, we
treat a time loop as a hypothetical physical system and investigate the
conditions for its quantum self-consistency.

Consider a particle that travels through a wormhole into its own past
and prevents its own motion. If the journey into the past occurs, the
younger particle must encounter the older one; if the collision prevents
the journey, the older particle should never have appeared. This paradox
shows that locally allowed classical motions need not combine into a
globally self-consistent history.

Classical models do not necessarily lead to this contradiction:
Echeverria, Klinkhammer, and Thorne
demonstrated self-consistent trajectories for a billiard ball
interacting with its own past version \cite{echeverria-1991}.
Classical self-consistency, self-interaction, and the causal structure of
wormhole spacetimes have also been studied more generally
\cite{friedman-1990,novikov-1992,friedman-morris-1997,arefeva-2008,fewster-1996,clement-2016}.
By contrast, we deliberately choose a one-dimensional geometry with an obstacle that is
impenetrable in the classical limit. The particles cannot simply ``pass
one another'': only specific channels of motion are available in the
classical problem.

Deutsch proposed imposing a self-consistency condition on the density
matrix of a system traveling along a closed timelike curve and showed
that finite-dimensional systems admit a fixed density matrix \cite{deutsch-1991}.
For infinite-dimensional quantum systems, the existence of a fixed
density matrix is guaranteed under various additional sufficient
conditions \cite{tumulka-weixler-2024}. This framework resolves classical logical
contradictions at the level of mixed quantum states, but its
interpretation remains controversial. Deutsch's many-worlds reading
is historically related to H.~Everett's relative-state formulation and
DeWitt's subsequent development of the many-worlds interpretation
\cite{everett-1957,dewitt-1970}. Dunlap \cite{dunlap-2016}, in particular,
argues that a physical reading of Deutsch's construction involves
additional ontological assumptions about a special kind of interacting
parallel worlds.

Other quantum treatments of time loops include path-integral
constructions \cite{politzer-1994} and postselected models
\cite{lloyd-2011}. For a comparison of circuit-based approaches, see
Allen \cite{allen-2014}. A.~Everett \cite{everett-2004} explored
self-consistency in a path-integral model and found that preserving
unitarity can make otherwise improbable events necessary -- the
so-called ``banana peel mechanism'' \cite{smith-1997}. Although such
approaches address quantum consistency, they do not by themselves
specify a single actual particle trajectory through the loop. This
leaves open the question of how an individual particle history can
satisfy the return condition.

We propose an extension of Deutsch's approach within de Broglie--Bohm
theory, in which, in addition to the quantum state, there is an actual
particle configuration guided by the corresponding wave field
\cite{de-broglie-1928,bohm-1952, drezet-amblard-2025,struyve-2010,durr-2004,beck-lazarovici-2025,aleshin-2024}.
We impose a self-consistency condition not only on the density
matrix but also on the particle's actual ontic configuration -- that is,
on its real trajectory in configuration space. The central question is
whether a self-consistent trajectory for an individual particle can
exist in the presence of a time loop.

For the self-prevention problem considered here, we answer in the
affirmative: an individual particle can have a self-consistent
trajectory in the presence of a time loop. The main result has two
parts: an explicit stationary density matrix in a two-channel Gaussian
model, and an analytically specified self-consistent ontic configuration
lying in the low-probability tunneling branch.

The article is organized as follows.
Section~\ref{general-framework-quantum-and-ontic-self-consistency}
introduces Deutsch's density-matrix condition, density-matrix guidance
in Bohmian mechanics, and the additional ontic closure requirement.
Section~\ref{one-dimensional-self-prevention-model} specifies the
two-channel return geometry, explains the classical self-prevention
paradox and its possible quantum resolution, and states the Hamiltonian
and approximations used in the calculation.
Section~\ref{self-consistent-density-matrix} establishes the leading-order
matching of the returned packet states and constructs the stationary
density matrix of the reduced two-channel model.
Section~\ref{self-consistent-bohmian-trajectory} derives the ontic fixed
point for equal and unequal packet widths and examines the convergence
of iterative methods based on the return map.
Section~\ref{discussion-and-conclusions} discusses the scope of the
results, their interpretation, and possible broader implications.

\section{General Framework: Quantum and Ontic
Self-Consistency}\label{general-framework-quantum-and-ontic-self-consistency}

We first state the two self-consistency conditions independently of the
particular collision model. The first is imposed on the quantum state,
as in Deutsch's approach; the second is an additional requirement on the
actual configuration in a Bohmian description. Both refer to the same
prescribed time-loop geometry.

\subsection{Deutsch's Density-Matrix
Condition}\label{deutschs-density-matrix-condition}

In Deutsch's general formulation \cite{deutsch-1991}, a chronology-respecting
system interacts with a system traversing a closed timelike curve (CTC).
For the self-interaction studied here, we denote the two input
subsystems by \(Y\) (younger) and \(O\) (older), respectively. The
symbols \(Y,O\) are labels for the two interacting quantum subsystems;
Deutsch's fixed-point construction does not in general require them to
be two stages of the same particle. Let \(\rho_Y\) be the prescribed
input state and \(\rho_O\) the unknown state returning from the future.

The collision is described by a unitary operator \(\mathcal V\) acting
on the two-particle system. In our geometry, the outgoing \emph{younger}
particle enters the return route and reappears as the incoming \emph{older}
particle. We therefore trace out the outgoing older subsystem and apply
the complete return channel \(\mathcal G\), which includes routing
through the wormhole mouths:

\begin{equation}\label{eq:deutsch-condition}
\rho_O
=\mathcal G\!\left(\operatorname{Tr}_{O}\!\left[
\mathcal V(\rho_Y\otimes\rho_O)\mathcal V^\dagger
\right]\right).
\end{equation}

Here \(\operatorname{Tr}_{O}\) traces over the \emph{outgoing} older
subsystem, which does not enter the return route; \(\mathcal G\)
identifies the outgoing younger state with the incoming older state.
This differs in notation from the usual Deutsch formula, where the
outgoing CTC subsystem is retained: an output exchange of subsystem
labels gives the same construction. The return channel may in particular
route spatially distinct output packets to different CTC exit regions.

Writing the resulting completely positive, trace-preserving map as
\(\mathcal D_{\rho_Y}\), we obtain the quantum consistency requirement

\begin{equation}\label{eq:quantum-consistency}
\boxed{
\rho_O=\mathcal D_{\rho_Y}(\rho_O).
}
\end{equation}

In finite-dimensional Hilbert spaces a fixed density matrix exists for
such a map \cite{deutsch-1991}. In infinite dimension, additional sufficient
conditions are required \cite{tumulka-weixler-2024}. This equation concerns the quantum
state; by itself, it does not specify one actual particle trajectory
that returns to its own initial configuration.

\subsection{Density-Matrix Guidance in Bohmian
Mechanics}\label{density-matrix-guidance-in-bohmian-mechanics}

In de Broglie--Bohm theory the quantum state is accompanied by an actual
configuration \(\mathbf Q=(Q_1,\ldots,Q_N)\) \cite{de-broglie-1928,bohm-1952}. A
density-matrix formulation \cite{durr-2005} allows the configuration to be
guided directly by a configuration-space density matrix \(W(q,q')\),
where \(q,q'\) are configuration arguments. For spinless particles, away
from nodes of \(W(\mathbf Q,\mathbf Q)\), the guiding equation reads

\begin{equation}\label{eq:density-matrix-guidance}
\dot Q_k=\frac{\hbar}{m_k}\operatorname{Im}
\left[
\frac{\nabla_{q_k}W(q,q')}{W(q,q')}
\right]_{q=q'=\mathbf Q}.
\end{equation}

Here \(m_k\) is the mass of particle \(k\); for internal spin indices
one uses the appropriate trace. The matrix \(W\) describes the relevant
full configuration, rather than being identified automatically with the
reduced \(\rho_O\). When spatial branches do not overlap at the
actual configuration, their contributions to the local velocity field
vanish, although branch interference during scattering can remain
important.

\subsection{Additional Ontic
Consistency}\label{additional-ontic-consistency}

Let \(\mathbf X_Y^{\mathrm{out}}\) be the actual outgoing configuration of
the younger subsystem entering a return route after interaction. Let
\(\mathbf X_O^{\mathrm{in}}\) denote the incoming older configuration
at the earlier matching surface. The geometric map \(\mathcal C\)
identifies these configurations, including spatial reorientation and
transport between the wormhole mouths. Our additional ontic
self-consistency condition, also referred to as trajectory closure, is

\begin{equation}\label{eq:ontic-consistency}
\boxed{
\mathbf X_O^{\mathrm{in}}=\mathcal C\!\left(\mathbf X_Y^{\mathrm{out}}\right).
}
\end{equation}

The outgoing configuration is determined by Bohmian motion under the
quantum state satisfying Eq.~\eqref{eq:quantum-consistency}. Thus the mathematical problem is to
find a joint solution of the density-matrix condition \eqref{eq:quantum-consistency} and
the ontic condition \eqref{eq:ontic-consistency}. A stationary density matrix need not
automatically imply a trajectory fixed point.

In the concrete model below, the CTC returns one and the same particle
to its own past. The general ontic requirement will reduce to equality
of two normalized internal coordinates, \(z_O=z_Y^{\mathrm{out}}\), on the
self-consistent tunneling branch. We will solve the quantum and ontic
conditions separately and then show their compatibility within the
stated Gaussian approximations.

\section{One-Dimensional Self-Prevention
Model}\label{one-dimensional-self-prevention-model}

\subsection{A Particle Traveling into Its Own
Past}\label{a-particle-traveling-into-its-own-past}

The particle initially moves from left to right through a region where
it may encounter its own older version. After the encounter, it follows
one of two routes, to the right or to the left; both lead to the wormhole
entrance. The wormhole returns it to a time before its initial passage
through the interaction region. The two routes connect to different
exit regions in the past, so the encounter determines where the older
version appears rather than whether the journey into the past occurs.
We prescribe these routes as boundary conditions of the effective model
without constructing a specific spacetime metric. The one-dimensional
restriction applies to the local collision: the particles cannot pass
one another on the side.

Two versions of the particle are therefore present near the interaction
region at the same time. We call the segment of its worldline
\emph{before entering} the wormhole the younger version \(Y\), and the
segment \emph{after emerging} from it in the past the older version
\(O\). In the calculation, these are two distinguishable degrees of
freedom in an effective two-particle model. They have equal mass \(m\);
their distinguishability can be associated with the particle's internal
clock.

An alternative formulation would let the younger particle enter the
wormhole only if it continues to the right: reflection by its older
version would then prevent the very journey that brought the older
version into its path. Although this expresses self-prevention more
directly, it requires quantum states with and without the older version
and transition rules between them. We use the two-route formulation
above.

To describe the collision quantitatively, we introduce a
repulsive two-particle interaction \(U(x_Y-x_O)\) that depends
on the separation between the two simultaneously present versions of the
particle. In the classical limit, we choose a potential that forms an
impenetrable barrier to their relative motion. Thus, an older
version that appears in the younger version's path does not merely
``forbid'' its motion by assumption: it physically reflects it.
For equal masses, this corresponds to an exchange of velocities in a
one-dimensional elastic collision. In the quantum problem, tunneling
through the same barrier is possible.

\subsubsection{Two Spatial Return
Channels}\label{two-spatial-return-channels}

If the younger version continues moving to the right after the
encounter, it enters the wormhole along the right-hand route and emerges
in the past to the right of itself. We denote this returning
version by \(O_A\), where \(A\) denotes the right-hand return sector. It moves toward the younger version and becomes
an obstacle in its path: the repulsive interaction
\(U(x_Y-x_O)\) creates a barrier that classically reflects the younger
particle to the left.

If, instead, the younger version is reflected to the left, it
enters the wormhole along the left-hand route and emerges in the past
to the left of itself; we denote this version by \(O_B\), where \(B\) denotes the left-hand return sector. The
exit time, position, and direction of motion are chosen so that, in this
case, the returning particle does not lie in the younger
version's path and does not prevent it from moving to the right.

Thus, the outcome of the collision with the particle's older
version determines where that same version emerges in the past:

\begin{equation}\label{eq:return-routing}
\mathsf T\longrightarrow O_A,
\qquad
\mathsf R\longrightarrow O_B.
\end{equation}

Here \(\mathsf T\) denotes transmission to the right and \(\mathsf R\)
reflection to the left. This rule connects the two physical routes
through the wormhole to its exit locations; it does not assume any
additional ``measurement'' or external decision about whether the
particle passes through the CTC.

Each route also defines a quantum return map: \(\mathcal U_{\mathsf T}\)
for the right-going transmitted particle and \(\mathcal U_{\mathsf R}\)
for the left-going reflected particle. These maps identify the outgoing
state of the younger version with the incoming state of the
older version at the corresponding CTC exit. They incorporate
translation, possible momentum reorientation, and propagation along the
respective route. In particular, they can modify the spatial width and
quadratic phase of a packet; they need not preserve its shape pointwise.
The maps and routing rule are model-defining boundary conditions, not
approximations.

\subsubsection{Classical Self-Prevention
Paradox and Its Possible Quantum Resolutions}\label{classical-self-prevention-paradox}

The routing rule produces an immediate contradiction in the classically
impenetrable regime. Suppose the older version appears to the
right (\(O_A\)). It then collides with the younger version and
stops its own earlier motion to the right: the classically
impenetrable repulsive barrier reflects the younger version, sending it
to the left. But this leftward route leads to the older version
appearing on the left, not the right:

\begin{equation}\label{eq:classical-right-to-left}
O_A\longrightarrow\mathsf R\longrightarrow O_B.
\end{equation}

Conversely, suppose the older version appears to the left
(\(O_B\)). It does not prevent the younger version from moving to the
right; the younger version enters the wormhole along the right-hand
route and must appear in the past on the right, not on the
left:

\begin{equation}\label{eq:classical-left-to-right}
O_B\longrightarrow\mathsf T\longrightarrow O_A.
\end{equation}

Therefore,

\begin{equation}\label{eq:classical-return-map}
O_B\longleftrightarrow O_A.
\end{equation}

Neither classical alternative reproduces its own initial conditions:
each returns the older particle to the opposite sector. Thus the chosen
one-dimensional interaction and two-channel geometry admit no
classically self-consistent individual history.

\begin{figure}[H]
  \centering
  \includegraphics[width=0.6\linewidth]{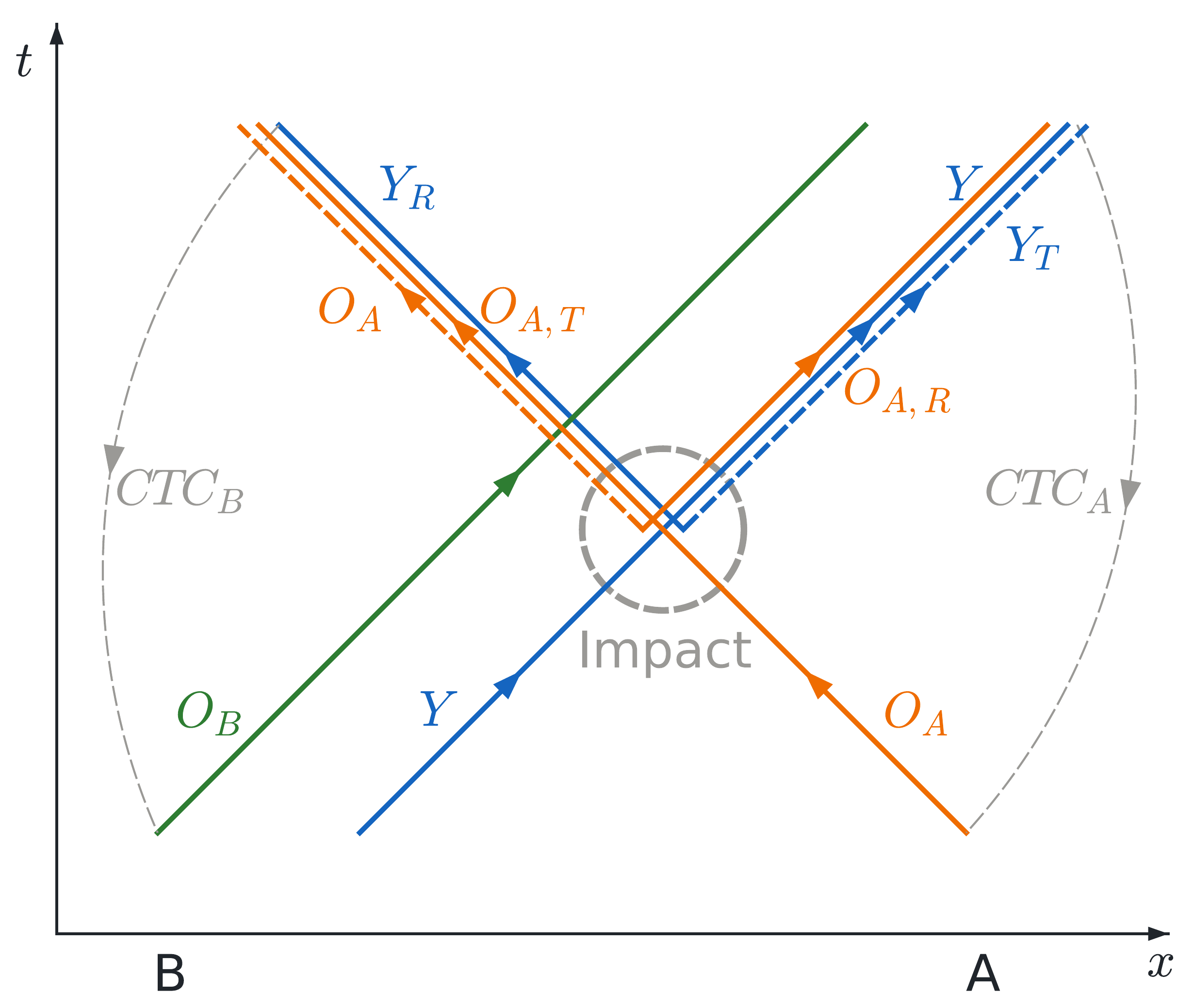}
  \caption{Schematic wave-packet trajectories through the encounter and
  the two return channels, including reflection and tunneling
  (time increases upward). Blue lines represent the younger particle
  \(Y\), with reflected and transmitted branches \(Y_R\) and \(Y_T\).
  The green line represents \(O_B\), which emerges on the left and does
  not interact with \(Y\). In this case, \(Y\) passes unchanged; the two
  labels \(Y\) mark its incoming and outgoing portions. Orange lines
  represent \(O_A\), which approaches from the right, and its outgoing
  branches after reflection or transmission in the relative motion. Dashed blue and orange lines indicate tunneling paths. Dashed grey curves indicate the return routes through the wormhole,
  connecting the outgoing younger particle to the incoming older
  particle at \(A\) (right) or \(B\) (left). The circle marks the
  interaction region.}
  \label{fig:spatial-return-channels}
\end{figure}

There are at least two possible quantum resolutions of this classical
self-prevention paradox (see Fig.~\ref{fig:spatial-return-channels}).
The first possibility involves many-worlds branching:
\begin{equation}\label{eq:many-worlds-return}
\begin{aligned}
O_A^1\longrightarrow\mathsf R\longrightarrow O_B^2, \\
O_B^2\longrightarrow\mathsf T\longrightarrow O_A^1,
\end{aligned}
\end{equation}
where the superscripts indicate different branches of the wave density matrix. This is the natural interpretation of the Deutsch model.

The second possibility, explored in this article, is a self-consistent
individual history without many-worlds branching. When the older version
is on the right, the
younger particle may tunnel through the repulsive interaction and take
the right-hand route:

\begin{equation}\label{eq:classical-contradiction}
O_A\xrightarrow{\text{tunneling}}\mathsf T
\xrightarrow{\mathrm{CTC}}O_A.
\end{equation}

This is the candidate individual consistent history. It must
satisfy the ontic condition of Section~\ref{additional-ontic-consistency},
while the full two-channel quantum state must simultaneously satisfy the
density-matrix condition of
Section~\ref{deutschs-density-matrix-condition}.

\subsection{Hamiltonian and
Scattering}\label{hamiltonian-and-scattering}

With the global routing specified by Eq.~\eqref{eq:return-routing} and
the return maps in Section~\ref{two-spatial-return-channels}, the local
collision is described by the effective two-particle Hamiltonian

\begin{equation}\label{eq:collision-hamiltonian}
H=-\frac{\hbar^2}{2m}
\left(\partial_{x_Y}^2+\partial_{x_O}^2\right)
+U(x_Y-x_O),
\end{equation}

where \(U\) is a repulsive interaction localized in the relative
coordinate and has a finite maximum \(U_{\max}\). Define

\begin{equation}\label{eq:collision-coordinates}
Q=\frac{x_Y+x_O}{2},
\qquad
\xi=x_Y-x_O;
\end{equation}

the Hamiltonian then becomes

\begin{equation}\label{eq:separated-hamiltonian}
H=-\frac{\hbar^2}{4m}\partial_Q^2
-\frac{\hbar^2}{m}\partial_\xi^2+U(\xi).
\end{equation}

The free evolution of the center of mass separates exactly at the
Hamiltonian level. Factorization of the two-particle wave function
itself in \(Q,\xi\) generally requires additional properties of the
incoming packets; in the symmetric case, it holds for two Gaussians with
matching complex widths.

We choose the relative-motion energy to be below the barrier threshold,
\(E_{\mathrm{rel}}<U_{\max}\), so that classical transmission is forbidden:
in the classical limit the barrier is impenetrable and reflects the
younger version. Let \(P\) be the quantum transmission probability when
the older version is on the right, with \(0<P<1\); in the regime
considered, this process is tunneling.

\subsection{Approximations}\label{approximations}

Our aim is to exhibit a self-consistent individual history in an
analytically tractable regime. This requires solving both the quantum
and ontic consistency conditions, but not a complete treatment of
arbitrary scattering states.

Even for simple interactions, such as rectangular or delta barriers,
an analytic solution of the stationary scattering problem does not
directly solve the full consistency problem. Wave-packet evolution
involves momentum integrals, and matching the outgoing and returning
states imposes additional integral relations. The return need not
preserve Gaussian form or factorization in center-of-mass and relative
coordinates.

Direct numerical evolution of the full two-particle density matrix
involves four spatial arguments and time, together with the return
consistency condition. Reduced numerical descriptions are possible but
lie beyond the present study.

We therefore adopt the assumptions below, which permit an explicit
analytic construction.

\assumption{ass:narrow-band}{Narrow-band Gaussian packets.} The incoming states of the
two versions are Gaussian and initially factorized in \(Y,O\). The
relative spectrum is centered at \(k_0\) with width \(\sigma_k\ll k_0\).
The spectrum of the transmitted branch is

\begin{equation}\label{eq:transmitted-spectrum}
\widetilde\chi_{\mathsf T}(k)=t(k)\widetilde\chi_{\mathrm{in}}(k),
\end{equation}

and that of the reflected branch is
\(\widetilde\chi_{\mathsf R}(k)=r_{\mathrm{sc}}(k)\widetilde\chi_{\mathrm{in}}(k)\), where \(t\) and
\(r_{\mathrm{sc}}\) are complex scattering amplitudes. Sufficient conditions
for weak distortion are

\begin{equation}\label{eq:weak-distortion}
\left|\frac{d\ln|t|}{dk}\right|_{k_0}\sigma_k\ll1,
\qquad
\left|\frac{d^2\ln t}{dk^2}\right|_{k_0}\sigma_k^2\ll1,
\end{equation}

with analogous conditions for \(r_{\mathrm{sc}}(k)\). The linear spectral
phase is incorporated into the spatial shift of the packet. Thus
\(P\simeq|t(k_0)|^2\); more precisely, \(P\) is the average of
\(|t(k)|^2\) over the relative momentum spectrum.

\assumption{ass:factorization}{Factorization of each outgoing branch.} For equal masses
and nearly constant scattering amplitudes, each outgoing branch
approximately factorizes in the single-particle coordinates. In the
transmitted channel, the younger particle retains its own packet; in the
reflected channel, it acquires the older particle's packet:

\begin{equation}\label{eq:branch-factorization}
\rho_{Y,\mathsf T}^{\mathrm{out}}\simeq\rho_Y,
\qquad
\rho_{Y,\mathsf R}^{\mathrm{out}}\simeq\rho_{O_A}.
\end{equation}

Within the narrow-band approximation, this also holds for different
incoming widths: equal widths are not required for factorization of
\emph{each individual outgoing branch}. The full superposition of
transmitted and reflected amplitudes is, in general, entangled.

\assumption{ass:branch-separation}{Branch separation.} The two spatial branches \(O_A,O_B\)
are orthogonal to a good approximation, and the outgoing channels
likewise separate after scattering. Their contributions to the density
matrix are then treated as distinct blocks. In the Bohmian account with
a density matrix, branches that have no support at the actual
configuration do not affect the local velocity field; interference
during the collision, before the outgoing waves separate, nevertheless
remains important \cite{durr-2005}. The correlation between the younger
particle's outgoing direction and the older particle's emergence
location acts as a record of the outcome (analogous to a which-channel
record). The partial trace removes inter-channel coherence in the
reduced description; it is not, by itself, an additional physical
measurement.

\assumption{ass:fast-collision}{Fast collision (used only for the ontic trajectory with
different widths, Section~\ref{different-widths-fast-collision-approximation}).} During the local interaction, the
normalized center-of-mass coordinate changes negligibly. At each
approximately fixed center-of-mass coordinate, the conditional
transmission probability is taken to be \(P\), independent of that
coordinate, and the normalized transmitted conditional packet is
approximated by a Gaussian. These assumptions make a one-dimensional
conditional quantile map possible; they do not follow solely from
narrow-band scattering.

\section{Self-Consistent Density
Matrix}\label{self-consistent-density-matrix}

We seek a fixed point of Deutsch's condition \eqref{eq:quantum-consistency} in the reduced,
block-diagonal two-channel model. A subtlety is that the
right-hand return channel receives contributions from two
different processes: unobstructed passage when the older particle is on
the left, and tunneling when it is on the right. We must justify
identifying their outgoing younger-particle states before combining the
contributions.

\subsection{Self-Consistency of Gaussian
States}\label{self-consistency-of-gaussian-states}

Let \(\rho_f\) be the normalized outgoing state of the younger particle
after unobstructed passage, and let \(\rho_t\) be its normalized reduced
state after tunneling, with the outgoing older particle traced out. Both
are defined at the same asymptotic matching surface, before propagation
through the right-hand CTC route. In general, \(\rho_f\) and \(\rho_t\)
need not be identical: their histories involve different correlations
with the older particle, and the scattering amplitude may introduce a
momentum-dependent phase.

The assumed narrow-band regime makes their leading-order
equality explicit. For equal masses, the transmission amplitude depends
on the relative wave number \(k=(p_Y-p_O)/(2\hbar)\), where
\(p_Y,p_O\) are the single-particle momenta. Over the incident
spectral width \(\sigma_k\), require not merely weak variation of
\(|t(k)|\) but also small variation of its phase relative to the freely
propagated packet:

\begin{equation}\label{eq:complex-amplitude-conditions}
\sigma_k\left|\frac{d\ln t}{dk}\right|_{k_0}\ll1,
\qquad
\sigma_k^2\left|\frac{d^2\ln t}{dk^2}\right|_{k_0}\ll1.
\end{equation}

The first condition includes the small-displacement requirement
\(\sigma_k|\varphi'(k_0)|\ll1\), where \(\varphi(k)=\arg t(k)\); a
constant phase is immaterial. We assume analogous weak distortion for
reflection, together with approximate factorization of the individual
outgoing branches as in Eq.~\eqref{eq:branch-factorization}. These conditions are fulfilled, for
example, in the contact-barrier limit \(U(\xi)=g\delta(\xi)\), \(g>0\),
where

\begin{equation}\label{eq:contact-barrier-amplitude}
t(k)=\frac{k}{k+i\gamma},\qquad
\gamma=\frac{mg}{2\hbar^2},\qquad
\left|\frac{d\ln t}{dk}\right|_{k_0}
=\frac{\sqrt{1-P}}{k_0}.
\end{equation}

Thus \(\sigma_k/k_0\ll1\) controls the relative variation of the
full complex amplitude, including its phase, for this
illustrative localized interaction, even if \(P\) is small. For wave
packets sufficiently concentrated near \(k_0\), the resulting normalized
younger-particle states satisfy \(\|\rho_t-\rho_f\|_1=O(\sigma_k/k_0)\)
in trace norm (with the return-route and free phases matched). A
sufficiently narrow, nonresonant finite barrier can be used
instead, provided the same spectral conditions are checked; spatial
localization alone does not guarantee them.

In the constant-amplitude limit the transmitted two-particle wave
function, after normalization and removal of free propagation, is
unchanged to leading order. Consequently, its younger-particle reduced
state is independent, to this order, of the detailed incoming state or
spatial translation of the older packet. We therefore have

\begin{equation}\label{eq:outgoing-state-identification}
\rho_f\simeq\rho_t\simeq\rho_Y.
\end{equation}

The right-hand return map \(\mathcal U_{\mathsf T}\) acts on either
outgoing younger state. The left-hand route applies after reflection;
for equal masses, the reflected younger particle approximately inherits
the older right-hand packet [Eq.~\eqref{eq:branch-factorization}]. The resulting
leading-order packet-shape conditions are

\begin{equation}\label{eq:gaussian-return-states}
\rho_{O_A}\simeq\mathcal U_{\mathsf T}(\rho_Y),
\qquad
\rho_{O_B}\simeq\mathcal U_{\mathsf R}(\rho_{O_A}).
\end{equation}

These conditions do not require equal widths: the return maps may
broaden the packets and generate quadratic phases, as long as the
incoming states remain narrow-band and the branchwise scattering
approximation applies. The well-separated spatial sectors \(O_A\) and
\(O_B\) are distinct from the two nearly coincident contributions
\(\rho_f\) and \(\rho_t\) within the right-hand sector. No
spatial separation between \(\rho_f\) and \(\rho_t\) is assumed.

\subsection{Stationary Density
Matrix}\label{stationary-density-matrix}

Let \(p_A,p_B\) be the probabilities of finding the incoming older
particle in the right and left spatial sectors, respectively, with
\(p_A+p_B=1\). A younger particle returns on the right either after
unobstructed passage from the left-older sector or after tunneling from
the right-older sector. These channel transitions are illustrated schematically in Fig.~\ref{fig:spatial-return-channels}. Therefore

\begin{equation}\label{eq:channel-evolution}
\begin{aligned}
p_A'&=p_B+Pp_A,\\
p_B'&=(1-P)p_A.
\end{aligned}
\end{equation}

For an approximately state-independent transmission probability \(P\),
stationarity gives \(p_A^*=1/(2-P)\) and \(p_B^*=1-p_A^*\).

To verify stationarity of the density operators, we first write the
unnormalized right-hand output before identifying \(\rho_t\) with
\(\rho_f\):

\begin{equation}\label{eq:right-return-mixture}
W_A'=p_B\mathcal U_{\mathsf T}(\rho_f)
      +p_AP\mathcal U_{\mathsf T}(\rho_t),
\end{equation}

where \(W_A'=p_A'\rho_{O_A}'\). The two terms correspond to unobstructed
passage and tunneling, respectively, with the outgoing older particle
traced out in each case. If \(\rho_f\) and \(\rho_t\) differ, the returned
state is a mixture of different packets.

Under the controlled leading-order identification
\(\rho_f\simeq\rho_t\simeq\rho_Y\), both contributions produce the same
returned state \(\rho_0:=\mathcal U_{\mathsf T}(\rho_Y)\). The resulting
candidate stationary operator is

\begin{equation}\label{eq:stationary-density-matrix}
\boxed{
\rho_O^*
\simeq\frac{1}{2-P}\rho_{O_A}
\oplus
\frac{1-P}{2-P}\rho_{O_B}.
}
\end{equation}

Here \(\oplus\) denotes orthogonal right and left spatial sectors. With \(W_A=p_A\rho_{O_A}\), \(W_B=p_B\rho_{O_B}\), the block
map becomes

\begin{equation}\label{eq:block-evolution}
\begin{aligned}
W_A'&\simeq
\bigl(\operatorname{Tr}W_B+P\operatorname{Tr}W_A\bigr)\rho_0,\\
W_B'&\simeq(1-P)\mathcal U_{\mathsf R}(W_A).
\end{aligned}
\end{equation}

Substituting the stationary weights and using
Eq.~\eqref{eq:gaussian-return-states} gives \(W_A'\simeq W_A\) and
\(W_B'\simeq W_B\) within this leading-order two-channel approximation.

The stationary weights remain valid when \(\rho_f\) and \(\rho_t\) differ,
provided \(P\) is state-independent. Replacing their mixture by a single
Gaussian packet requires the complex-amplitude conditions above;
otherwise, each spatial sector may contain several packets. The ontic
analysis in Section~\ref{self-consistent-bohmian-trajectory} uses the
leading-order Gaussian state, since trace-norm closeness alone does not
control Bohmian trajectories in remote Gaussian tails.

\section{Self-Consistent Bohmian
Trajectory}\label{self-consistent-bohmian-trajectory}

The density-matrix solution of Section~\ref{self-consistent-density-matrix} fixes the statistical CTC
state. We next impose the additional ontic requirement \eqref{eq:ontic-consistency} on the branch
in which the older particle appears on the right and the younger
particle tunnels to the right. We first treat equal widths, then allow
arbitrary positive width ratios compatible with the Gaussian and
fast-collision approximations. No explicit propagation time or spreading
law is needed for the latter result.

\subsection{Equal Gaussian Widths}\label{equal-gaussian-widths}

\subsubsection{Relative-Coordinate
Quantile}\label{relative-coordinate-quantile}

First consider the symmetric case in which the packets of the
younger particle and the older particle on the right have matching
complex Gaussian width parameters. Let \(\sigma>0\) denote their common
spatial standard deviation before the collision. Their actual positions
are

\begin{equation}\label{eq:equal-width-positions}
X_Y=\bar X_Y+\sigma z_Y,
\qquad
X_O=\bar X_O+\sigma z_O.
\end{equation}

Introduce the normalized variables

\begin{equation}\label{eq:symmetric-coordinates}
z_Q=\frac{z_Y+z_O}{\sqrt2},
\qquad
r=\frac{z_Y-z_O}{\sqrt2}.
\end{equation}

For independent incoming Gaussians, both variables are standard normally
distributed. Denote the standard normal cumulative distribution function
by

\begin{equation}\label{eq:normal-cdf}
\Phi(a)=\frac12\left[1+\operatorname{erf}\left(\frac a{\sqrt2}\right)\right].
\end{equation}
Its density is \(\phi(a)=e^{-a^2/2}/\sqrt{2\pi}\).

In one-dimensional Bohmian relative motion, trajectories do not cross
\cite{oriols-1996}. This ordering underlies the quantile description of
Bohmian trajectories \cite{brandt-1998} and the identification of incoming
positions leading to transmission \cite{sanz-2011}.
With the chosen orientation, the transmitted branch corresponds to
incoming quantiles \(q_{\mathrm{in}}>1-P\), where \(q_{\mathrm{in}}=\Phi(r)\).
Once the branches have separated, one-dimensional ordering fixes the
quantile \(q_{\mathsf T}\) of the actual trajectory within the normalized
transmitted packet. The additional assumption that this packet remains
Gaussian relates that quantile to its normalized outgoing relative
coordinate \(r_{\mathsf T}\):

\begin{equation}\label{eq:transmitted-relative-coordinate}
q_{\mathsf T}=\Phi(r_{\mathsf T})=\frac{\Phi(r)-1+P}{P}.
\end{equation}

Here \(\Phi^{-1}\), used below, denotes the inverse cumulative
distribution function.

\subsubsection{Ontic Fixed Point}\label{ontic-fixed-point}

In the symmetric factorized problem, the normalized center-of-mass
coordinate is preserved, so

\begin{equation}\label{eq:outgoing-younger-equal-widths}
z_Y^{\mathrm{out}}=\frac{z_Q+r_{\mathsf T}}{\sqrt2}.
\end{equation}

By assumption, the CTC map preserves the normalized internal coordinate of
the Gaussian packet, even under free spreading. Therefore, trajectory
closure requires

\begin{equation}\label{eq:ontic-closure}
z_O=z_Y^{\mathrm{out}}.
\end{equation}

\begin{figure}[H]
  \centering
  \includegraphics[width=0.82\linewidth]{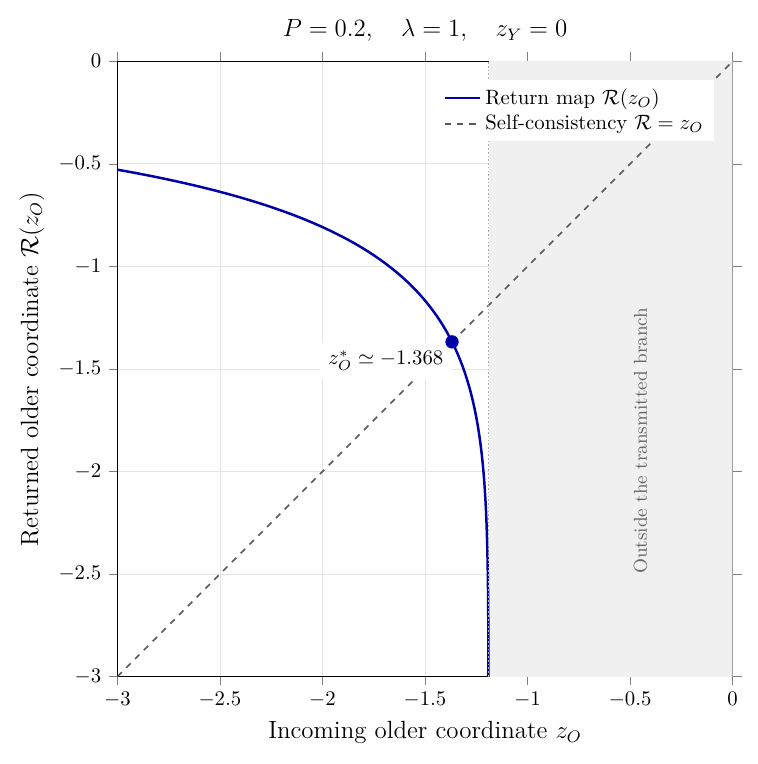}
  \caption{Ontic return map \eqref{eq:symmetric-return-map} in the
  symmetric Gaussian model with equal packet widths, for \(P=0.2\)
  and \(z_Y=0\). The solid curve gives
  the returned older coordinate \(\mathcal R(z_O)=z_Y^{\mathrm{out}}\)
  as a function of the incoming older coordinate \(z_O\). Its
  intersection with the dashed diagonal \(\mathcal R=z_O\) gives the
  self-consistent history, with \(z_O^*\simeq-1.368\).
  Equal axis scales make the diagonal a \(45^\circ\) line. The map is
  defined on the transmitted branch,
  \(z_O<z_Y-\sqrt2\,\Phi^{-1}(1-P)\);
  the shaded region lies outside this domain.}
  \label{fig:ontic-return-map}
\end{figure}

To visualize this condition, define the return map
\(\mathcal R(z_O)=z_Y^{\mathrm{out}}\) at fixed \(P\) and \(z_Y\).
For equal packet widths, Eqs.~\eqref{eq:symmetric-coordinates} and
\eqref{eq:transmitted-relative-coordinate} give the explicit map
\begin{equation}\label{eq:symmetric-return-map}
\boxed{
\mathcal R(z_O)=\frac{z_Y+z_O}{2}
+\frac{1}{\sqrt2}\,\Phi^{-1}\!\left[
\frac{\Phi\!\left((z_Y-z_O)/\sqrt2\right)-1+P}{P}
\right].
}
\end{equation}
It is defined on the transmitted branch,
\(z_O<z_Y-\sqrt2\,\Phi^{-1}(1-P)\).
The outgoing younger particle becomes the incoming older particle after
return through the CTC. An individual history therefore closes at an
intersection of \(\mathcal R(z_O)\) with the diagonal
\(\mathcal R=z_O\), as illustrated in Figure~\ref{fig:ontic-return-map}.

Equations~\eqref{eq:symmetric-coordinates} and \eqref{eq:ontic-closure}
imply \(r_{\mathsf T}=-r\). Substitution into
Eq.~\eqref{eq:transmitted-relative-coordinate}, using
\(\Phi(-r)=1-\Phi(r)\), gives

\begin{equation}\label{eq:symmetric-fixed-point}
r_*=\Phi^{-1}\left(\frac{1}{1+P}\right).
\end{equation}

For a given internal coordinate \(z_Y\) of the younger particle, we
obtain

\begin{equation}\label{eq:symmetric-older-coordinate}
\boxed{
z_O^*=z_Y-\sqrt2\,\Phi^{-1}\left(\frac{1}{1+P}\right).
}
\end{equation}

The resulting fixed point lies specifically in the tunneling channel:

\begin{equation}\label{eq:tunneling-membership}
\Phi(r_*)-(1-P)=\frac{P^2}{1+P}>0.
\end{equation}

As \(P\to0^+\), the coordinate \(r_*\) moves into the right tail of the
normal distribution, while \(z_O^*\to-\infty\) for fixed \(z_Y\). In the
strict classical limit \(P=0\), no finite tunneling fixed point exists.

\begin{figure}[H]
  \centering
  \includegraphics[width=0.95\linewidth]{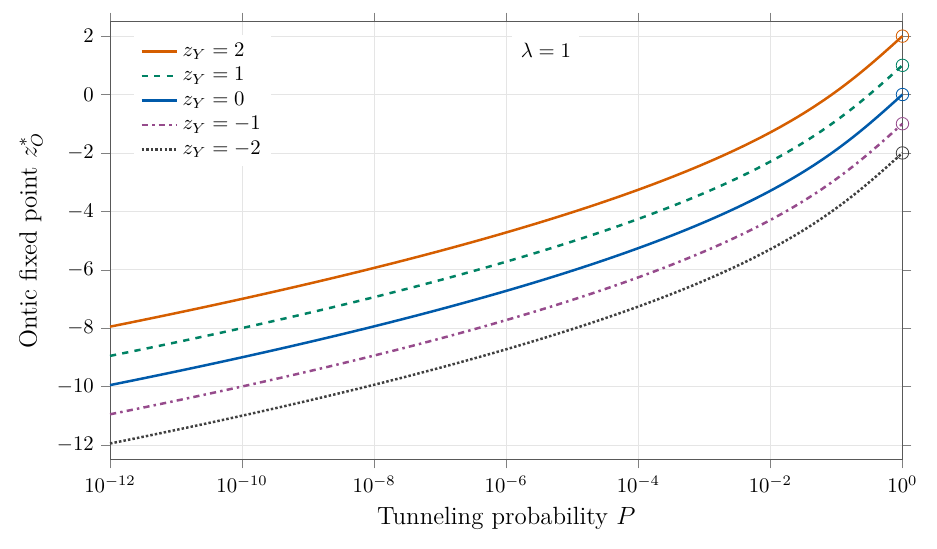}
  \caption{Ontic fixed point \(z_O^*\) as a function of the tunneling
  probability \(P\), evaluated from
  Eq.~\eqref{eq:symmetric-older-coordinate} in the symmetric Gaussian
  model with equal packet widths (\(\lambda=1\)) for
  \(z_Y=-2,-1,0,1,2\). The probability axis is logarithmic.
  As \(P\to0^+\), \(z_O^*\to-\infty\); no finite fixed point exists at
  \(P=0\). Open circles mark the limiting values \(z_O^*=z_Y\) as
  \(P\to1^-\). The curves differ only by the vertical shift \(z_Y\).}
  \label{fig:ontic-fixed-point-probability}
\end{figure}

Figure~\ref{fig:ontic-fixed-point-probability} shows the dependence of
this ontic fixed point on the tunneling probability for equal packet
widths and several fixed values of \(z_Y\). The \(z_Y=0\) curve
includes the fixed point illustrated in Figure~\ref{fig:ontic-return-map}
at \(P=0.2\).

\subsubsection{Convergence of Direct Iterations of the Return
Map}\label{convergence-of-return-map-iterations}

The return map \eqref{eq:symmetric-return-map} also defines a numerical
procedure for finding an ontic fixed point. Here we examine its local
convergence in the equal-width Gaussian model, with \(P\in(0,1)\) and
\(z_Y\) held fixed. The map is smooth on its transmitted domain, and
its fixed point lies strictly inside that domain by
Eq.~\eqref{eq:tunneling-membership}.

Consider the direct iteration
\begin{equation}\label{eq:direct-return-iteration}
z_O^{(n+1)}=\mathcal R\!\left(z_O^{(n)}\right).
\end{equation}
Differentiating Eq.~\eqref{eq:symmetric-return-map} and using
\(r_{\mathsf T}=-r_*\) at the fixed point gives its
linear error multiplier:
\begin{equation}\label{eq:direct-iteration-multiplier}
\mathcal R'(z_O^*)=-\frac{1-P}{2P}.
\end{equation}
Thus direct iteration is locally convergent for \(P>1/3\), with errors
alternating in sign. For \(0<P<1/3\), the fixed point is locally
repelling under this iteration. In particular, for the example
\(P=0.2\) in Figure~\ref{fig:ontic-return-map}, the linear error
multiplier is \(-2\).

At \(P=1/3\), the multiplier is \(-1\), so linearization alone is
inconclusive. With \(\delta_n=z_O^{(n)}-z_O^*\), expanding two
successive iterations yields
\begin{equation}\label{eq:critical-return-iteration}
\delta_{n+2}=\delta_n+2(1-r_*^2)\delta_n^3+O(\delta_n^4),
\qquad r_*=\Phi^{-1}(3/4)\simeq0.67449.
\end{equation}
The positive cubic coefficient implies weak local repulsion at this
boundary value as well.

Convergence can be restored by mixing the returned coordinate with the
previous iterate, also called under-relaxation:
\begin{equation}\label{eq:relaxed-return-iteration}
z_O^{(n+1)}=(1-\omega)z_O^{(n)}
+\omega\mathcal R\!\left(z_O^{(n)}\right),
\qquad 0<\omega\le1.
\end{equation}
For any nonzero \(\omega\), this modification preserves the fixed-point
equation. Its local error multiplier is
\begin{equation}\label{eq:relaxed-iteration-multiplier}
q_\omega=1-\omega\frac{1+P}{2P}.
\end{equation}
The condition \(|q_\omega|<1\) gives the sufficient local convergence
range
\begin{equation}\label{eq:relaxation-convergence-range}
0<\omega<\frac{4P}{1+P},
\qquad 0<\omega\le1.
\end{equation}
For every \(P\in(0,1)\), choosing \(\omega=2P/(1+P)\) sets
\(q_\omega=0\) and gives at least quadratic local convergence
(\(\omega=1/3\) for \(P=0.2\)).

More generally, for a smooth scalar return map with derivative
\(d=\mathcal R'(z_O^*)\), mixing gives the local criterion
\(|1-\omega+\omega d|<1\). This permits a suitable mixing parameter to
stabilize an oscillatory direct iteration when \(d<0\), although the
appropriate parameter depends on the map. The explicit bounds above
apply to the equal-width model. All convergence statements are local:
a sufficiently close initial iterate remains within a contracting
neighborhood in the transmitted domain, whereas distant iterates may
leave that domain.

\subsection{Different Widths: Fast-Collision
Approximation}\label{different-widths-fast-collision-approximation}

\subsubsection{Normalized Variables}\label{normalized-variables}

Let \(\lambda=\sigma_O/\sigma_Y>0\) denote the ratio of the spatial
widths of the older right-hand and younger packets at the collision. The return maps may generate such a difference and a nonzero
quadratic phase. At the start of the collision,

\begin{equation}\label{eq:unequal-width-positions}
X_Y=\bar X_Y+\sigma_Yz_Y,
\qquad
X_O=\bar X_O+\sigma_Oz_O.
\end{equation}

Let \(\Sigma=\sqrt{\sigma_Y^2+\sigma_O^2}\) and define the orthogonal
normalized coordinates

\begin{equation}\label{eq:normalized-coordinates}
\begin{aligned}
s&=\frac{\sigma_Yz_Y+\sigma_Oz_O}{\Sigma},\\
u&=\frac{\sigma_Oz_Y-\sigma_Yz_O}{\Sigma}.
\end{aligned}
\end{equation}

Here \(Q,\xi\) are evaluated at the actual Bohmian positions
\(x_Y=X_Y\), \(x_O=X_O\). The quantity
\(s=2(Q-\bar Q)/\Sigma\) is the normalized center-of-mass coordinate.
The variable \(u\) is the normalized relative coordinate
at fixed \(s\); indeed,

\begin{equation}\label{eq:relative-coordinate-decomposition}
\xi-\bar\xi=\frac{\sigma_Y^2-\sigma_O^2}{\Sigma}s+
\frac{2\sigma_Y\sigma_O}{\Sigma}u.
\end{equation}

For factorized incoming Gaussians, \(s\) and \(u\) are independent and
standard normally distributed.

\subsubsection{Conditional Quantile and Closure
Equation}\label{conditional-quantile-and-closure-equation}

Under the fast-collision and conditional Gaussian assumptions (\ref{ass:fast-collision}), the
normalized transmitted relative coordinate is mapped according to

\begin{equation}\label{eq:conditional-quantile}
u_{\mathsf T}=
\Phi^{-1}\left(\frac{\Phi(u)-1+P}{P}\right).
\end{equation}

Within the same approximation, the incoming and outgoing internal
coordinates of the younger particle are related by

\begin{equation}\label{eq:outgoing-younger-coordinate}
z_Y^{\mathrm{out}}=\frac{s+\lambda u_{\mathsf T}}{\sqrt{1+\lambda^2}}.
\end{equation}

Since free Gaussian propagation along the CTC preserves the normalized
internal coordinate of an individual particle, ontic self-consistency
again requires \(z_O=z_Y^{\mathrm{out}}\). Substituting the coordinate
transformation \eqref{eq:normalized-coordinates} into this condition and
using Eq.~\eqref{eq:conditional-quantile} gives a scalar closure equation
with coefficients

\begin{equation}\label{eq:closure-coefficients}
\alpha=\lambda-1+\frac1\lambda,
\qquad
\beta=\frac{(\lambda-1)\sqrt{1+\lambda^2}}{\lambda}.
\end{equation}

The resulting equation is

\begin{equation}\label{eq:closure-equation}
\boxed{
\Phi(u)-P\Phi(\beta z_Y-\alpha u)=1-P.
}
\end{equation}

Once the solution \(u_*\) is found, the older coordinate is given by

\begin{equation}\label{eq:older-fixed-coordinate}
\boxed{
z_O^*=\lambda z_Y-\sqrt{1+\lambda^2}\,u_*.
}
\end{equation}

\subsubsection{Existence of a Root in the Gaussian
Map}\label{existence-of-a-root-in-the-gaussian-map}

For any \(\lambda>0\), we have
\(\alpha=\lambda+\lambda^{-1}-1\ge1\). Define

\begin{equation}\label{eq:closure-function}
f(u)=\Phi(u)-1+P-P\Phi(\beta z_Y-\alpha u).
\end{equation}

Differentiation gives

\begin{equation}\label{eq:closure-monotonicity}
f'(u)=\phi(u)+P\alpha\phi(\beta z_Y-\alpha u)>0.
\end{equation}

As \(u\to-\infty\), \(f(u)\to-1\), whereas as \(u\to+\infty\),
\(f(u)\to P\). Therefore, for any \(P\in(0,1)\), \(\lambda>0\), and
fixed \(z_Y\), Eq.~\eqref{eq:closure-equation} has a unique finite root in this
approximate map. Equation~\eqref{eq:closure-equation} also implies
\(\Phi(u_*)>1-P\), so the root lies in the transmitted channel.

For \(\lambda=1\), we have \(\alpha=1\) and \(\beta=0\), so the general
fixed-point equation \eqref{eq:closure-equation} reduces to the symmetric result \eqref{eq:symmetric-fixed-point}--\eqref{eq:symmetric-older-coordinate}.
Thus, unequal widths do not remove the root within the
fast-collision Gaussian map.

\section{Discussion and
Conclusions}\label{discussion-and-conclusions}

In this work, we considered the self-prevention paradox in Deutsch's
time-loop model, supplemented with a Bohmian description of an
individual particle. In the classical version of the model, no
self-consistent trajectory exists: the spatial return channels lead to a
contradiction. In the quantum description, a stationary density matrix
is found; in the Bohmian description, there is a self-consistent
individual history in which the younger particle tunnels through its
interaction with the older one. For any nonzero tunneling probability,
the resulting history must pass through this channel, even though such
an event may be extremely unlikely in ordinary scattering. At zero
tunneling probability, this type of individual solution disappears. This
realizes the ``banana peel mechanism'' \cite{smith-1997}: the global requirement
of self-consistency selects a locally unlikely event as necessary for a
contradiction-free history.

The key methodological step is the choice of approximations that make
both the quantum and ontic self-consistency conditions analytically
tractable. Within this approximation, we explicitly construct a
stationary density matrix and a self-consistent individual history,
using the quantile description of Bohmian motion for the latter. This
provides an analytic demonstration of the tunneling resolution of the
paradox.

The result is established for the prescribed one-dimensional,
two-channel model under the assumptions of Section~\ref{approximations}.
Whether such individual histories persist beyond these approximations
remains an open question. A relativistic extension would additionally
require a specification of the spacetime structure used to define
Bohmian configurations, including the role of spacelike foliations
\cite{durr-2014}.

These results raise a broader question: how do interpretations of
quantum mechanics affect physical reality in the presence of closed
timelike curves? In ordinary circumstances, without closed timelike
curves, standard interpretations of quantum mechanics agree on
observable predictions. Deutsch's time-loop model, however, requires an
additional self-consistency condition, making the question of what
counts as a physical history consequential. In a many-worlds reading,
the solution is described in terms of interacting parallel branches of
history. In a Bohmian reading, one can speak of the trajectory of an
individual particle; in the model considered here, that trajectory
passes through a low-probability tunneling channel. Thus, the presence
of closed timelike curves makes interpretation more than a way of
talking about the quantum state: it affects the physical account of how
the paradox is resolved and what kind of self-consistent history is
realized.

This motif can be cautiously extended to a macroscopic time-travel
paradox. In discussions of the ``grandfather paradox,'' the banana peel
mechanism means that unlikely accidents may stop the time traveler
\cite{smith-1997}. If, however, the self-prevention is caused by a reliable,
deterministic mechanism, it is harder to see how that mechanism could
fail in a classical description. In a many-worlds reading of Deutsch's
approach, the contradiction is resolved through interactions between
parallel branches of history. In a Bohmian reading, one can imagine a
different scenario: individual components of the mechanism tunnel,
causing it to fail. This is only a possible analogy with the result of
our model, not a conclusion about the feasibility of closed timelike
curves or macroscopic tunneling.

\section*{Use of Artificial Intelligence}

OpenAI's ChatGPT was used to support analytical derivations,
mathematical calculations, consistency checks, and the preparation and
editing of the manuscript. The author reviewed and verified all
AI-assisted content, independently reproduced the central calculations,
and takes full responsibility for the scientific content and conclusions
of this work.

\section*{Notation and Abbreviations}

\begin{description}[font=\normalfont,style=nextline,leftmargin=2em]
\item[CTC]
  \emph{closed timelike curve}.
\item[\(Y,O\)]
  \emph{younger} and \emph{older} interacting subsystems; in the model,
  the two versions of the particle. \(O_A\) is the right-hand older
  sector and \(O_B\) is the left-hand older sector.
\item[\(\mathsf T,\mathsf R\)]
  \emph{transmission} and \emph{reflection} outcomes of the younger
  particle; \(P\) -- the tunneling probability when the older particle
  occupies sector \(A\).
\item[\(x_Y,x_O\)]
  coordinate arguments of the two-particle wave function;
  \(X_Y,X_O\) -- actual Bohmian positions;
  \(\bar X_Y,\bar X_O\) -- packet centers.
\item[\(m,\hbar,H,U\)]
  particle mass, reduced Planck constant, Hamiltonian and repulsive
  interaction potential, respectively.
\item[\(Q,\xi\)]
  center-of-mass and relative coordinates;
  \(\bar Q,\bar\xi\) -- their mean positions;
  \(\Psi\) -- the two-particle wave function.
\item[\(k,k_0,\sigma_k\)]
  relative wave number, its central value and spectral width;
  \(p_Y,p_O\) -- single-particle momenta;
  \(t,r_{\mathrm{sc}}\) -- complex transmission and reflection
  amplitudes; \(\widetilde\chi_{\mathrm{in}}\) -- incoming spectral amplitude.
\item[\(\rho_Y,\rho_O\)]
  prescribed younger and self-consistent older density matrices;
  \(\rho_{O_A},\rho_{O_B}\) -- normalized older states in the two
  spatial sectors; \(\mathcal D_{\rho_Y}\) -- quantum map for a cycle.
\item[\(\rho_f,\rho_t,\rho_0\)]
  normalized outgoing younger states after unobstructed passage and
  tunneling, and the common returned state
  \(\rho_0=\mathcal U_{\mathsf T}(\rho_Y)\), respectively.
\item[\(\mathcal U_{\mathsf T},\mathcal U_{\mathsf R}\)]
  return channels associated with the transmitted and reflected
  outcomes; \(p_A,p_B\) -- sector weights;
  \(W_A,W_B\) -- their unnormalized density-matrix blocks.
\item[\(\mathcal V,\mathcal G\)]
  collision unitary and complete quantum return channel;
  \(\operatorname{Tr}_{O}\) -- partial trace over the outgoing older
  subsystem.
\item[\(\sigma_Y,\sigma_O,\sigma\)]
  positive spatial standard deviations of the younger and older
  packets and their common value in the symmetric limit;
  \(\Sigma=\sqrt{\sigma_Y^2+\sigma_O^2}\) -- combined width;
  \(\lambda=\sigma_O/\sigma_Y\) -- width ratio.
\item[\(z_Y,z_O,z_Q,r,s,u\)]
  normalized internal coordinates, including symmetric center-of-mass
  and relative coordinates \(z_Q,r\), and unequal-width coordinates
  \(s,u\); \(r_{\mathsf T},u_{\mathsf T}\) -- transmitted relative
  coordinates.
\item[\(q_{\mathrm{in}},q_{\mathsf T}\)]
  incoming relative-coordinate quantile and the quantile within the
  normalized transmitted packet.
\item[\(\mathcal R\)]
  ontic return map \(\mathcal R(z_O)=z_Y^{\mathrm{out}}\), at fixed
  \(P,z_Y\), in the equal-width model.
\item[\(\omega,q_\omega\)]
  mixing parameter and local error multiplier of the mixed return-map
  iteration.
\item[\(\Phi,\phi,\operatorname{erf}\)]
  standard normal cumulative distribution, its density and the error
  function; \(\Phi^{-1}\) -- the inverse normal cumulative distribution.
\item[\(\alpha,\beta\)]
  scalar coefficients in Eq.~\eqref{eq:closure-coefficients};
  \(f\) -- auxiliary function; superscripts \(\mathrm{in}\) and
  \(\mathrm{out}\) -- incoming and outgoing quantities;
  \(*\) -- value at the self-consistent solution.
\item[\(W(q,q')\)]
  full configuration-space density matrix in the guiding equation;
  \(\mathbf Q=(Q_1,\ldots,Q_N)\) -- actual configuration;
  \(m_k\) -- mass of particle \(k\);
  \(\nabla_{q_k}\) -- its configuration gradient;
  \(\mathcal C\) -- geometric return map;
  \(\mathbf X_Y^{\mathrm{out}},\mathbf X_O^{\mathrm{in}}\) -- actual outgoing
  younger and incoming older configurations.
\end{description}

\end{document}